\RequirePackage{lineno}

\documentclass[aps,prl,twocolumn,showpacs,superscriptaddress,letterpaper,amsmath,amssymb]{revtex4}

\usepackage{graphicx}  
\usepackage{dcolumn}   
\usepackage{bm}        
\usepackage{amssymb}   
\usepackage{feynmf}
\usepackage{slashed}
\def\gsim{\lower0.5ex\hbox{$\:\buildrel >\over\sim\:$}}
\def\lsim{\lower0.5ex\hbox{$\:\buildrel <\over\sim\:$}}

\def \n{\noindent}

\let\n=\nu
\let\t=\tau

\newcommand{\be}{\begin{equation}}
\newcommand{\ee}{\end{equation}}
\newcommand{\bea}{\begin{eqnarray}}
\newcommand{\eea}{\end{eqnarray}}

\newcommand{\nbox}{{\,\lower0.9pt\vbox{\hrule \hbox{\vrule height 0.2 cm
\hskip 0.2 cm \vrule height 0.2 cm}\hrule}\,}}

\def\missET {{\not\!\! E_T}}

\begin{document}

\thispagestyle{empty}
\vspace*{-3.5cm}

\vspace{0.5in}


\bibliographystyle{apsrev}

\title{{Search for a two-Higgs-boson doublet\\ using a simplified model in {\bf $p\bar{p}$} collisions at {\bf $\sqrt{s}=1.96$ TeV}}}

\affiliation{Institute of Physics, Academia Sinica, Taipei, Taiwan 11529, Republic of China}
\affiliation{Argonne National Laboratory, Argonne, Illinois 60439, USA}
\affiliation{University of Athens, 157 71 Athens, Greece}
\affiliation{Institut de Fisica d'Altes Energies, ICREA, Universitat Autonoma de Barcelona, E-08193, Bellaterra (Barcelona), Spain}
\affiliation{Baylor University, Waco, Texas 76798, USA}
\affiliation{Istituto Nazionale di Fisica Nucleare Bologna, $^{ee}$University of Bologna, I-40127 Bologna, Italy}
\affiliation{University of California, Davis, Davis, California 95616, USA}
\affiliation{University of California, Irvine, Irvine, California 92697, USA}
\affiliation{University of California, Los Angeles, Los Angeles, California 90024, USA}
\affiliation{Instituto de Fisica de Cantabria, CSIC-University of Cantabria, 39005 Santander, Spain}
\affiliation{Carnegie Mellon University, Pittsburgh, Pennsylvania 15213, USA}
\affiliation{Enrico Fermi Institute, University of Chicago, Chicago, Illinois 60637, USA}
\affiliation{Comenius University, 842 48 Bratislava, Slovakia; Institute of Experimental Physics, 040 01 Kosice, Slovakia}
\affiliation{Joint Institute for Nuclear Research, RU-141980 Dubna, Russia}
\affiliation{Duke University, Durham, North Carolina 27708, USA}
\affiliation{Fermi National Accelerator Laboratory, Batavia, Illinois 60510, USA}
\affiliation{University of Florida, Gainesville, Florida 32611, USA}
\affiliation{Laboratori Nazionali di Frascati, Istituto Nazionale di Fisica Nucleare, I-00044 Frascati, Italy}
\affiliation{University of Geneva, CH-1211 Geneva 4, Switzerland}
\affiliation{Glasgow University, Glasgow G12 8QQ, United Kingdom}
\affiliation{Harvard University, Cambridge, Massachusetts 02138, USA}
\affiliation{Division of High Energy Physics, Department of Physics, University of Helsinki and Helsinki Institute of Physics, FIN-00014, Helsinki, Finland}
\affiliation{University of Illinois, Urbana, Illinois 61801, USA}
\affiliation{The Johns Hopkins University, Baltimore, Maryland 21218, USA}
\affiliation{Institut f\"{u}r Experimentelle Kernphysik, Karlsruhe Institute of Technology, D-76131 Karlsruhe, Germany}
\affiliation{Center for High Energy Physics: Kyungpook National University, Daegu 702-701, Korea; Seoul National University, Seoul 151-742, Korea; Sungkyunkwan University, Suwon 440-746, Korea; Korea Institute of Science and Technology Information, Daejeon 305-806, Korea; Chonnam National University, Gwangju 500-757, Korea; Chonbuk National University, Jeonju 561-756, Korea}
\affiliation{Ernest Orlando Lawrence Berkeley National Laboratory, Berkeley, California 94720, USA}
\affiliation{University of Liverpool, Liverpool L69 7ZE, United Kingdom}
\affiliation{University College London, London WC1E 6BT, United Kingdom}
\affiliation{Centro de Investigaciones Energeticas Medioambientales y Tecnologicas, E-28040 Madrid, Spain}
\affiliation{Massachusetts Institute of Technology, Cambridge, Massachusetts 02139, USA}
\affiliation{Institute of Particle Physics: McGill University, Montr\'{e}al, Qu\'{e}bec, Canada H3A~2T8; Simon Fraser University, Burnaby, British Columbia, Canada V5A~1S6; University of Toronto, Toronto, Ontario, Canada M5S~1A7; and TRIUMF, Vancouver, British Columbia, Canada V6T~2A3}
\affiliation{University of Michigan, Ann Arbor, Michigan 48109, USA}
\affiliation{Michigan State University, East Lansing, Michigan 48824, USA}
\affiliation{Institution for Theoretical and Experimental Physics, ITEP, Moscow 117259, Russia}
\affiliation{University of New Mexico, Albuquerque, New Mexico 87131, USA}
\affiliation{The Ohio State University, Columbus, Ohio 43210, USA}
\affiliation{Okayama University, Okayama 700-8530, Japan}
\affiliation{Osaka City University, Osaka 588, Japan}
\affiliation{University of Oxford, Oxford OX1 3RH, United Kingdom}
\affiliation{Istituto Nazionale di Fisica Nucleare, Sezione di Padova-Trento, $^{ff}$University of Padova, I-35131 Padova, Italy}
\affiliation{University of Pennsylvania, Philadelphia, Pennsylvania 19104, USA}
\affiliation{Istituto Nazionale di Fisica Nucleare Pisa, $^{gg}$University of Pisa, $^{hh}$University of Siena and $^{ii}$Scuola Normale Superiore, I-56127 Pisa, Italy}
\affiliation{University of Pittsburgh, Pittsburgh, Pennsylvania 15260, USA}
\affiliation{Purdue University, West Lafayette, Indiana 47907, USA}
\affiliation{University of Rochester, Rochester, New York 14627, USA}
\affiliation{The Rockefeller University, New York, New York 10065, USA}
\affiliation{Istituto Nazionale di Fisica Nucleare, Sezione di Roma 1, $^{jj}$Sapienza Universit\`{a} di Roma, I-00185 Roma, Italy}
\affiliation{Rutgers University, Piscataway, New Jersey 08855, USA}
\affiliation{Texas A\&M University, College Station, Texas 77843, USA}
\affiliation{Istituto Nazionale di Fisica Nucleare Trieste/Udine, I-34100 Trieste, $^{kk}$University of Udine, I-33100 Udine, Italy}
\affiliation{University of Tsukuba, Tsukuba, Ibaraki 305, Japan}
\affiliation{Tufts University, Medford, Massachusetts 02155, USA}
\affiliation{University of Virginia, Charlottesville, Virginia 22906, USA}
\affiliation{Waseda University, Tokyo 169, Japan}
\affiliation{Wayne State University, Detroit, Michigan 48201, USA}
\affiliation{University of Wisconsin, Madison, Wisconsin 53706, USA}
\affiliation{Yale University, New Haven, Connecticut 06520, USA}

\author{T.~Aaltonen}
\affiliation{Division of High Energy Physics, Department of Physics,
University of Helsinki and Helsinki Institute of Physics, FIN-00014,
Helsinki, Finland}
\author{J.~Adelman}
\affiliation{Yale University, New Haven, Connecticut 06520, USA}
\author{B.~\'{A}lvarez~Gonz\'{a}lez$^z$}
\affiliation{Instituto de Fisica de Cantabria, CSIC-University of Cantabria, 39005 Santander, Spain}
\author{S.~Amerio}
\affiliation{Istituto Nazionale di Fisica Nucleare, Sezione di Padova-Trento, $^{ff}$University of Padova, I-35131 Padova, Italy}
\author{D.~Amidei}
\affiliation{University of Michigan, Ann Arbor, Michigan 48109, USA}
\author{A.~Anastassov$^x$}
\affiliation{Fermi National Accelerator Laboratory, Batavia, Illinois 60510, USA}
\author{A.~Annovi}
\affiliation{Laboratori Nazionali di Frascati, Istituto Nazionale di Fisica Nucleare, I-00044 Frascati, Italy}
\author{J.~Antos}
\affiliation{Comenius University, 842 48 Bratislava, Slovakia; Institute of Experimental Physics, 040 01 Kosice, Slovakia}
\author{G.~Apollinari}
\affiliation{Fermi National Accelerator Laboratory, Batavia, Illinois 60510, USA}
\author{J.A.~Appel}
\affiliation{Fermi National Accelerator Laboratory, Batavia, Illinois 60510, USA}
% ...

\collaboration{CDF Collaboration\footnote{With visitors from
$^a$Istituto Nazionale di Fisica Nucleare, Sezione di Cagliari, 09042 Monserrato (Cagliari), Italy,
$^b$University of CA Irvine, Irvine, CA 92697, USA,
$^c$University of CA Santa Barbara, Santa Barbara, CA 93106, USA,
$^d$University of CA Santa Cruz, Santa Cruz, CA 95064, USA,
$^e$Institute of Physics, Academy of Sciences of the Czech Republic, Czech Republic,
$^f$CERN, CH-1211 Geneva, Switzerland,
$^g$Cornell University, Ithaca, NY 14853, USA,
$^h$University of Cyprus, Nicosia CY-1678, Cyprus,
$^i$Office of Science, U.S. Department of Energy, Washington, DC 20585, USA,
$^j$University College Dublin, Dublin 4, Ireland,
$^k$ETH, 8092 Zurich, Switzerland,
$^l$University of Fukui, Fukui City, Fukui Prefecture, Japan 910-0017,
$^m$Universidad Iberoamericana, Mexico D.F., Mexico,
$^n$University of Iowa, Iowa City, IA 52242, USA,
$^o$Kinki University, Higashi-Osaka City, Japan 577-8502,
$^p$Kansas State University, Manhattan, KS 66506, USA,
$^q$Ewha Womans University, Seoul, 120-750, Korea,
$^r$University of Manchester, Manchester M13 9PL, United Kingdom,
$^s$Queen Mary, University of London, London, E1 4NS, United Kingdom,
$^t$University of Melbourne, Victoria 3010, Australia,
$^u$Muons, Inc., Batavia, IL 60510, USA,
$^v$Nagasaki Institute of Applied Science, Nagasaki, Japan,
$^w$National Research Nuclear University, Moscow, Russia,
$^x$Northwestern University, Evanston, IL 60208, USA,
$^y$University of Notre Dame, Notre Dame, IN 46556, USA,
$^z$Universidad de Oviedo, E-33007 Oviedo, Spain,
$^{aa}$CNRS-IN2P3, Paris, F-75205 France,
$^{bb}$Texas Tech University, Lubbock, TX 79609, USA,
$^{cc}$Universidad Tecnica Federico Santa Maria, 110v Valparaiso, Chile,
$^{dd}$Yarmouk University, Irbid 211-63, Jordan,
}}
\noaffiliation

\begin{abstract}
 We present a search for new particles in an extension to the standard
 model that includes a heavy Higgs boson ($H^0$), a lighter charged Higgs
 boson ($H^\pm$), 
 and an even-lighter Higgs boson $h^0$, with decays leading to  a $W$-boson
 pair and a bottom-antibottom quark pair in the 
final state. We use events with exactly one lepton, missing
 transverse momentum, and at least four jets in data corresponding to an integrated
 luminosity of 8.7 fb$^{-1}$ collected by the CDF II detector in
 proton-antiproton collisions at $\sqrt{s}=1.96$ TeV. We find the data to be consistent with  
 standard model predictions and report the results in terms of a
 simplified Higgs-cascade-decay model,
 setting 95\% confidence level upper limits on the product of
 cross-section and branching fraction from 1.3 pb to 15 fb  as a function of $H^0$ and $H^\pm$
 masses for $m_h^{0}=126$ GeV$/c^2$.
\end{abstract}



\maketitle

\date{\today}

The study of the mechanism of electroweak-symmetry breaking is one of
the major thrusts of the experimental high-energy-physics
program. Following the discovery of a Higgs-like boson at
ATLAS~\cite{atlashiggs} and CMS~\cite{cmshiggs} with a mass of approximately 126 GeV$/c^2$ and
complementary evidence from CDF and D0~\cite{cdfd0higgs},
the most pressing question is whether this state is in fact the Higgs
boson of the  standard model (SM), part of an extended Higgs
sector (such as that of the minimal supersymmetric standard model,
MSSM \cite{mssm}),
a composite Higgs \cite{comphiggs},
or a completely different
particle with Higgs-like couplings
(such as a radion in warped extra dimensions \cite{radionHiggs}
or a dilaton \cite{dilatonHiggs}).

We search for particles in an extension to the standard model that includes a
light neutral Higgs boson, $h^0$, with mass $m_{h^0}=126$ GeV$/c^2$.   Rather than assuming a particular theoretical framework (such as the
MSSM), we follow a phenomenological approach, using a general
two-Higgs doublet model as a convenient simplified
model \cite{simpmodels}, which contains a heavy charged Higgs boson $H^\pm$ and a heavier
neutral state $H^0$. In this approach, the search for a number of specific final states that have the strongest couplings to Higgs particles is motivated~\cite{spin0res,spinWZtbsimpmodel}.
The  final state of a $W$-boson pair ($WW$) is enhanced by $WW$ scattering
in models where the Higgs sector is strongly coupled \cite{strongww}. This signal has been the subject of much detailed investigation
\cite{wwscatteringLHC}. The phenomenology of resonant production of the
final states $Zh^0$ \cite{ZH} and $W^+ W^- Z$ \cite{WWZ}
has also been investigated.

In this letter, we focus on the final state $W^+ W^- b\bar{b}$~\cite{ourpaper},
which can have a large production rate from the process $gg \to H^0$ 
followed by $H^0 \to H^\pm W^\mp$ with 
$H^\pm \to W^\pm h^0 \to W^\pm b\bar{b}$.   The $W^+W^-b\bar{b}$ final
state is also the final state of top-quark pair production, and has been extensively
studied. However, no search for Higgs-boson cascades as described
here has been reported previously,  though searches have been performed for charged
Higgs bosons in top-quark decays $t\rightarrow H^\pm b$~\cite{chiggs1,chiggs2,chiggs3}. 

We analyze a data sample corresponding to an integrated
luminosity of 8.7$\pm 0.5$ fb$^{-1}$ recorded by the CDF II
detector~\cite{cdf}, a general purpose detector designed to study
$p\overline{p}$ collisions at $\sqrt{s}=1.96$ TeV in the Fermilab Tevatron collider. 
The CDF tracking system consists of a silicon microstrip tracker and a
drift chamber that are immersed in a 1.4 T axial magnetic field~\cite{trackingsystems}. 
Projective-tower-geometry electromagnetic and hadronic calorimeters surrounding
the tracking system measure particle energies, with muon detection
provided by  additional drift chambers
located outside the calorimeters.

The signature of $H^0 \to W^\mp H^\pm \to W^-W^+ h^0 \to W^-W^+ b\bar{b}$  is a charged lepton ($e$ or $\mu$), missing
transverse momentum, two jets arising from $b$ quarks, and two
additional jets from a $W$-boson hadronic decay. 
Events are selected online (triggered) by the requirement of an
electron ($e$) or muon ($\mu$) candidate~\cite{leptonid} with
transverse momentum $p_T$~\cite{coordinates} greater than $18$ GeV$/c$.
After trigger selection, events are retained if the electron or muon candidate has a pseudorapidity $|\eta| < 1.1$~\cite{coordinates}, $p_T > 20$ GeV$/c$, and satisfies the standard CDF identification and
isolation requirements~\cite{leptonid}. 
We reconstruct jets in the
calorimeter using the {\sc jetclu}~\cite{jetclu} algorithm with a
clustering radius of 0.4 in $\eta-\phi$ space. The jets are calibrated using
the techniques outlined in Ref.~\cite{jes}. At least four jets are
required, each with transverse energy $E_T > 15$ GeV and $|\eta| < 2.4$. Missing transverse momentum~\cite{metdef}
is reconstructed using 
calorimeter and muon information~\cite{leptonid}; in the $W^+W^-b\bar{b}$ experimental
signature,
the missing transverse momentum is mostly due to the neutrino from the
leptonically-decaying $W$ boson. We require $\missET>20$ GeV$/c$.
Since such a signal would yield two jets originating from
$b$ quarks,
we require (with minimal loss of efficiency) evidence of decay of a
$b$ hadron in at least one jet. This requirement, called $b$-tagging,
makes use of the {\sc secvtx} algorithm, which identifies jets from
$b$ quarks via their secondary vertices~\cite{secvtx}.

We model the production of  $H^0$ bosons with $m_{H^0} = 325$--$1100$
GeV$/c^2$ and subsequent decays $H^0\to W^\mp H^\pm$ with
$m_{H^\pm}=225$--$600$ GeV$/c^2$ and decays $H^\pm\to W^\pm h^0$ with $m_{h^0}=126$
GeV$/c^2$, all with {\sc madgraph}~\cite{madgraph}. Additional radiation, hadronization, and
showering are described by {\sc pythia}~\cite{pythia}. The detector
response for all simulated samples is modeled by the {\sc geant}-based CDF II detector simulation~\cite{cdfsimul}.

The dominant SM background to this signature is top-quark
pair production. We model this background using
{\sc pythia} with a top-quark mass $m_t = 172.5$ GeV/$c^2$~\cite{topmass}. We
normalize the $t\bar{t}$ background to the theoretical calculation at
next-to-next-to-leading order (NNLO) in the strong interaction
coupling constant, $\alpha_s$~\cite{nloa}.  In addition,
events generated by a next-to-leading order program, {\sc mc@nlo}~\cite{mcatnlo} are used in estimating an uncertainty
	    in modeling the radiation of an additional jet.

 The second largest SM background process is the associated production 
of a $W$ boson and jets. Samples of  $W$-boson+jets events with
light- and heavy-flavor ($b,c$) quark
jets are generated using {\sc alpgen}~\cite{alpgen},
and interfaced with a parton-shower model 
from {\sc pythia}.  The $W$-boson+jets samples are normalized to the
measured $W$-boson-production cross section, 
with an additional multiplicative factor for the relative contribution of heavy- 
and light-flavor jets, following Ref.~\cite{secvtx}.

 Backgrounds due
to production of a $Z$ boson with additional jets, where the second
lepton from the $Z$-boson decay is not reconstructed, are small
compared to the $W$-boson background and are modeled using events
generated with {\sc alpgen} interfaced to the parton-shower model 
from {\sc pythia}. The multi-jet
background, in which a jet is
misreconstructed as a lepton, is modeled using events triggered on
jets and
 normalized
to a background-dominated region at low missing transverse momentum
where the multi-jet background is large.

The SM backgrounds due to production of single top quarks and pairs of
vector bosons  are modeled
using {\sc madgraph} interfaced with {\sc pythia} parton-shower models and {\sc pythia}, respectively, and normalized to next-to-leading-order
cross sections~\cite{ewknlo,ewknlo2}. 

\begin{figure}[tp]
\includegraphics[width=.7\linewidth]{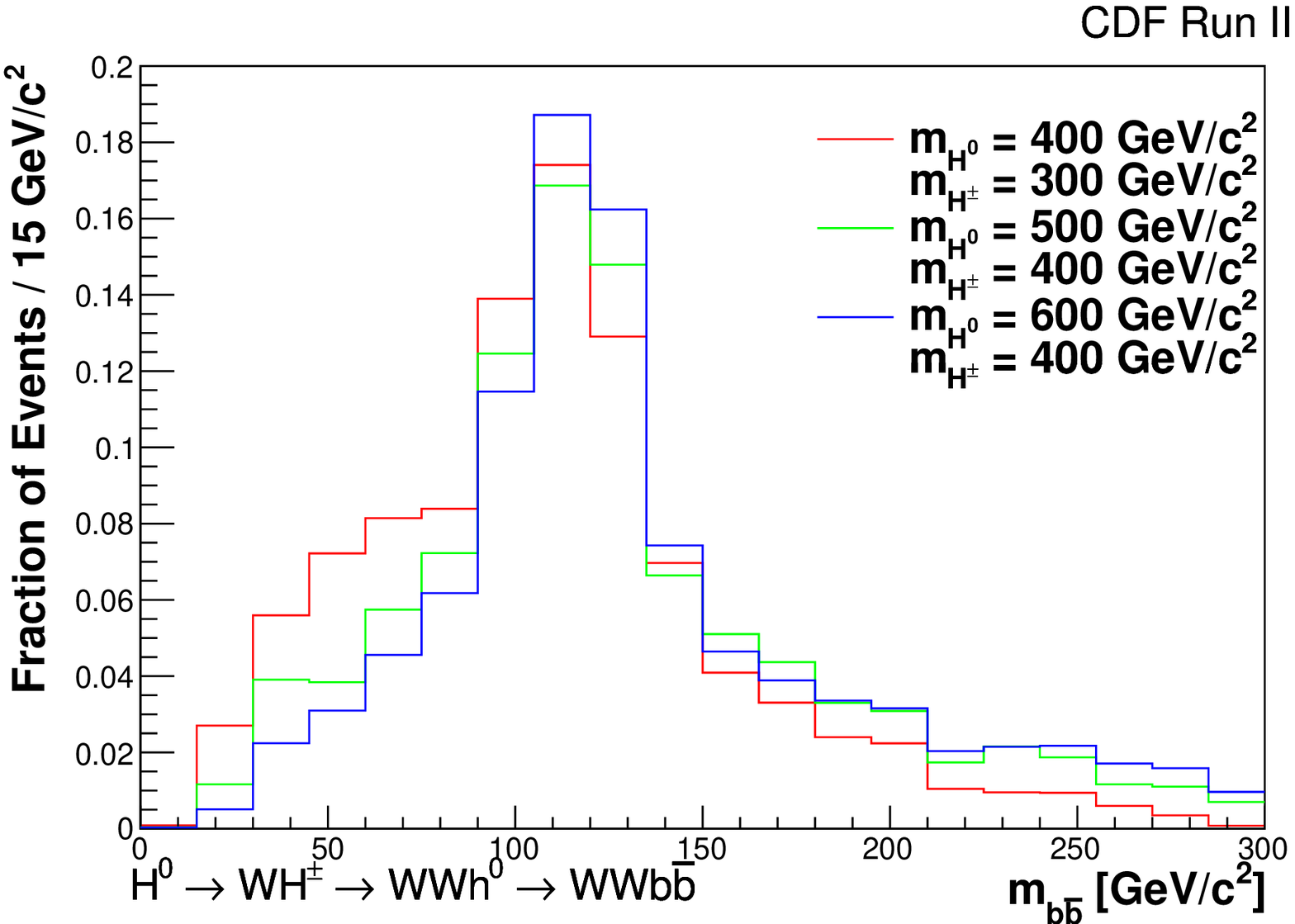}
\includegraphics[width=.7\linewidth]{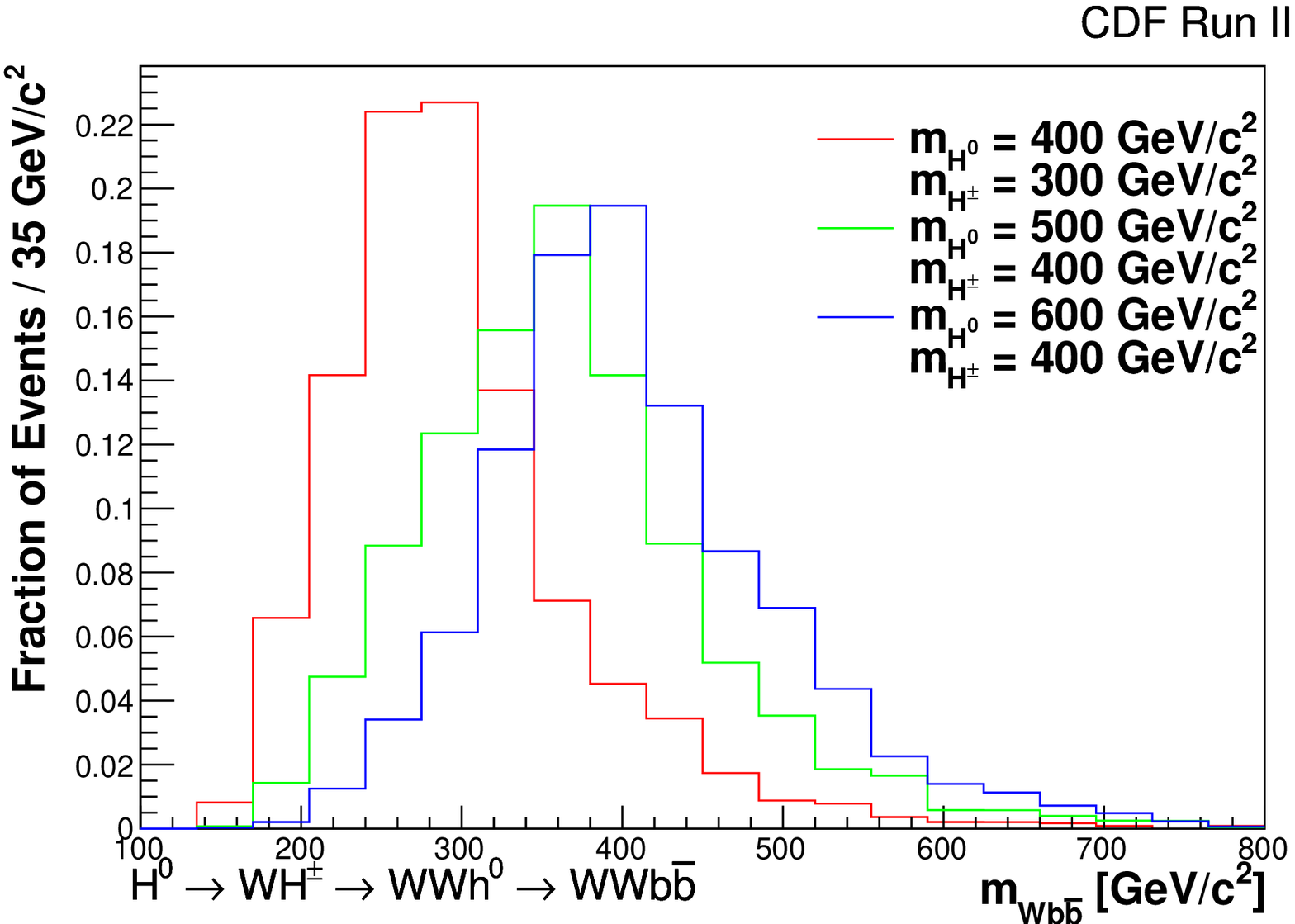}
\includegraphics[width=.7\linewidth]{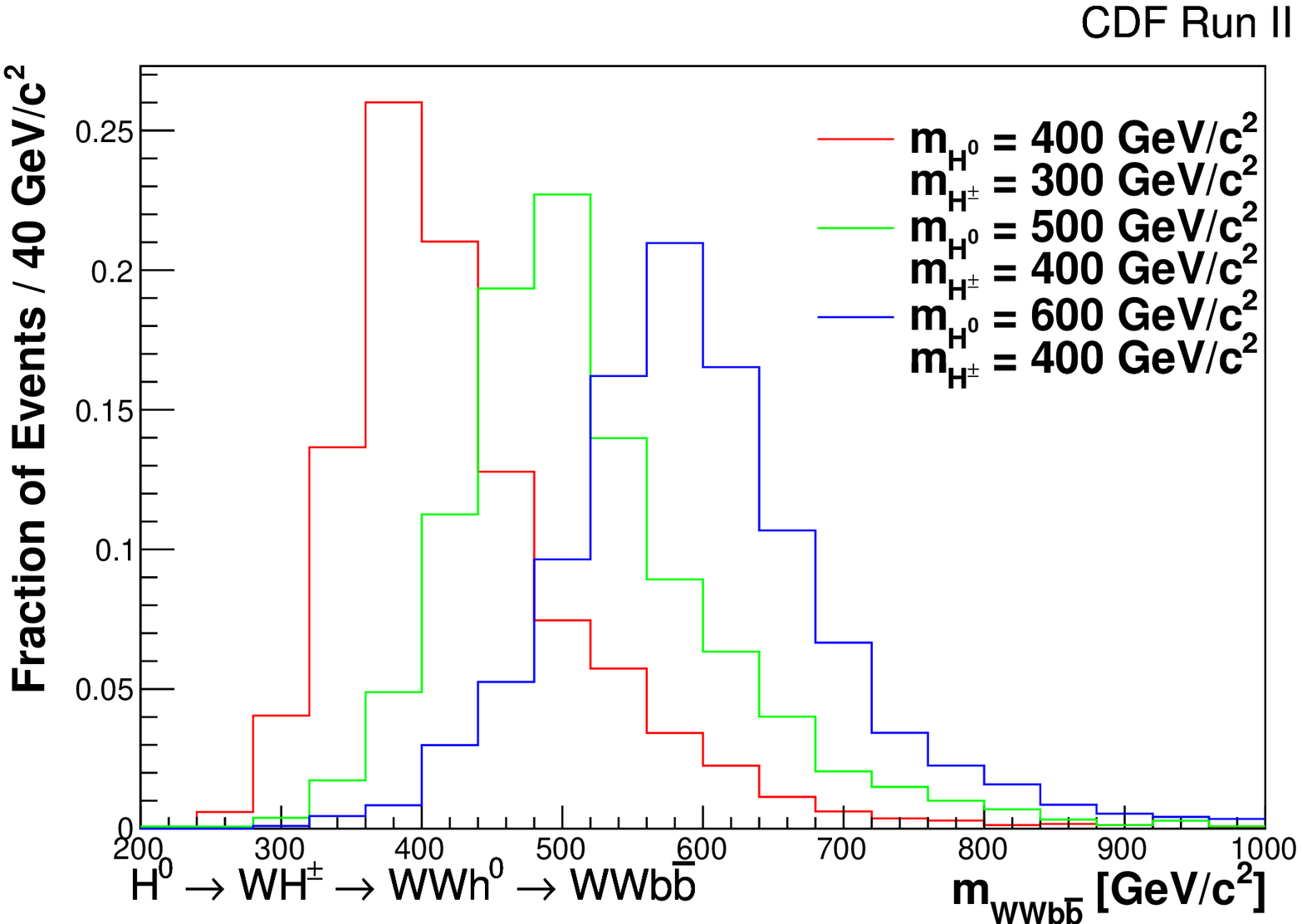}
\caption{Distribution of reconstructed Higgs-boson masses in simulated
  events. Top: $m_{h^0}=126$ GeV/$c^2$ reconstructed as $m_{bb}$, center: $m_{H^\pm}$ as
  $m_{Wbb}$, and bottom: $m_{H^0}$ as $m_{WWbb}$. }
\label{fig:reco}
\end{figure}

The Higgs-boson candidate mass reconstruction begins with
identification of the
leptonically-decaying $W$ boson, assuming the missing transverse
momentum is due to the resulting neutrino.  Of the multiple solutions for the neutrino pseudorapidity, we use the smallest value that yields the reconstructed $W$ mass closest to the known value. The hadronically-decaying
$W$ boson is identified as the pair of jets that yield the reconstructed dijet mass closest to the known $W$ mass, excluding jets with a $b$-tag. If  fewer than
two jets without $b$-tags are present, the same procedure is used but modified to
include the $b$-tagged jets.  The light  $h^0$
is reconstructed from the remaining $b$-tagged jets. If 
fewer than two $b$-tagged jets remain, the jet or jets with largest 
transverse momentum not associated with the hadronic $W$-boson decay are used
instead, without significant loss of mass resolution.  Figure~\ref{fig:reco} shows distributions of the
reconstructed mass for several choices of Higgs masses.

We enhance the signal-to-background ratio through requirements on
the mass of the $W^+W^-b\bar{b}$ and $W^\pm b\bar{b}$ systems, and search for an excess of events above expectations
from backgrounds in event distributions versus the mass of the $b\bar{b}$
system ($h^0\rightarrow b\bar{b}$).   Backgrounds have broad, smoothly decreasing distributions  while a signal
 would be reconstructed near the Higgs-boson mass.

We consider several sources of systematic uncertainty on the predicted background
rates and distributions, as well as on the expectations for a
signal. Each systematic uncertainty affects the expected sensitivity
to a signal, expressed
as an expected cross-section upper limit in the no-signal
assumption. The dominant systematic uncertainty is the jet-energy-scale  uncertainty~\cite{jes}, followed by theoretical uncertainties on the
cross sections of the background processes.  To probe the description
of additional jets, we compare our nominal $t\bar{t}$ model to one
generated by {\sc mc@nlo} and take the full difference as a systematic
uncertainty.  We also consider systematic uncertainties associated with the description of
initial- and final-state radiation~\cite{rad}, uncertainties in the efficiency of
reconstructing leptons and identifying $b$-quark jets, and uncertainties in the contribution
from multiple  interactions.  In addition, we consider
a variation of the $Q^2$ scale of $W$-boson+jet events in {\sc algpen}. In each case, we treat
the unknown underlying quantity as a nuisance parameter.  Except in
the case of the normalization uncertainty, which affects only the
overall rates, for each source of uncertainty we measure the distortion of the
 $m_{b\bar{b}}$ spectrum for positive and negative fluctuations of the
 underlying quantity.  Table~\ref{tab:syst} lists the
 contributions of each of these sources of systematic uncertainty to the
yields.

\begin{table}[!h]
\caption{Contributions to the systematic uncertainty on the expected numbers of events for the two main background processes, the total background yield,  and
  an example $500$ GeV$/c^2$ Higgs-boson signal with an assumed
  total cross section of 1 pb. }
\begin{ruledtabular}
\begin{tabular}{lcccc}
 Process & $t\bar{t}$ & $W$-boson+jets &  Total bg.& Higgs \\ \hline
Predicted yield & 229 & 43& 294 & 341\\  \hline
Jet energy scale & $23\%$ & - & $17\%$ & $12\%$ \\
Radiation & $3\%$ & -  & $2\%$ & $8\%$\\
$Q^2$ scale& - & $18\%$  &$3\%$  & - \\
Mult.\ interactions & $1\%$ & $6\%$ & $2\%$ & - \\
$t\bar{t}$ generator & $5\%$ &  -  & $4\%$ & -\\ 
Normalization & $10\%$ & $30\%$ & $16\%$ & - \\
\hline
Total\ syst.\ uncert.& $26\%$ &  $35\%$ & $24\%$ & $15\%$\\ 
\end{tabular}
\end{ruledtabular}
\label{tab:syst}
\end{table}

\begin{figure}
\includegraphics[width=0.9\linewidth]{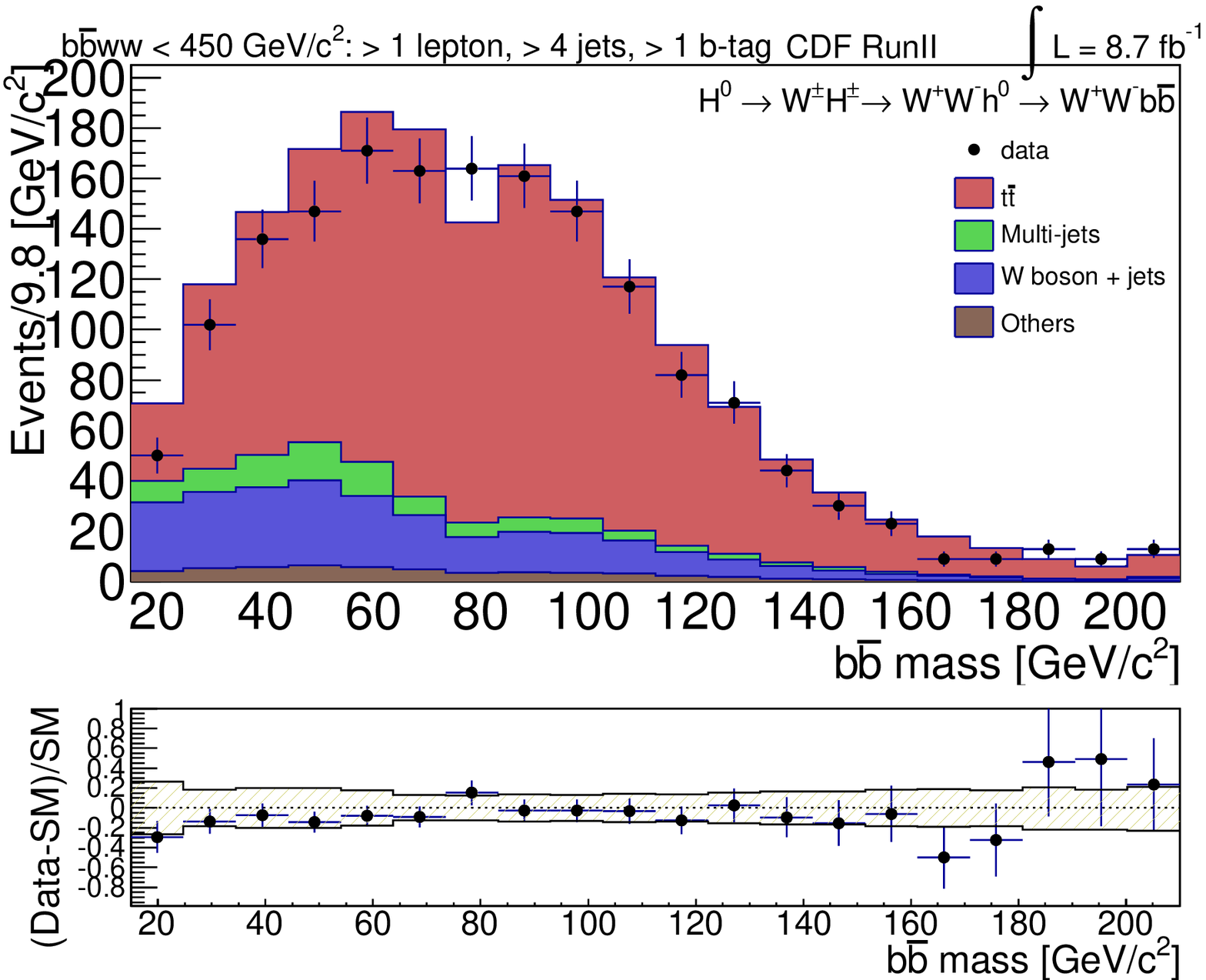}\\
\includegraphics[width=0.9\linewidth]{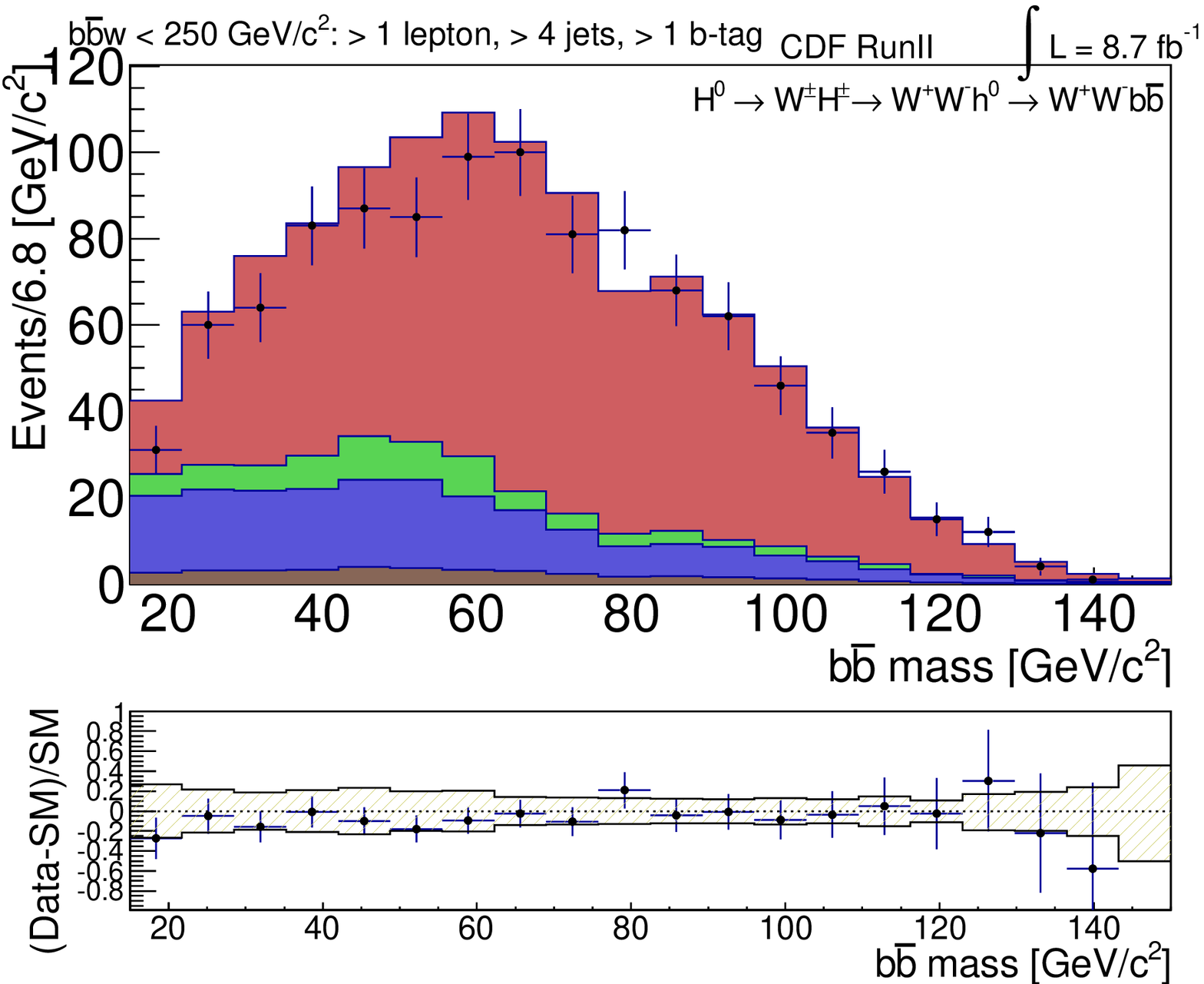}\\
\caption{Distribution of events versus reconstructed $b\bar{b}$  invariant mass ($m_{b\bar{b}}$) for observed data and expected backgrounds in two 
control regions. Top,  control region consisting events with at least four jets, exactly
zero $b$-tags and $m_{WWbb}<450$  GeV$/c^2$. 
Bottom, control region consisting of events with at least four jets and
$m_{Wbb}<250$ GeV$/c^2$.
 The lower panels give the relative
  difference between the observed and expected distributions; the
  hatched areas show the combined statistical and systematic
  uncertainties of the expected background. The small dip near 80 GeV$/c^2$ is mainly due to the
$W$-boson mass
reconstruction.}
\label{fig:ctrl_region}
\end{figure}

We validate our modeling of the SM backgrounds in four
background-dominated control regions. Each control region preserves the one lepton and at least four jet requirements with additional
requirements per region. Events in the first region are used to study
the $W^+W^-b\bar{b}$ and $W^\pm b\bar{b}$ mass reconstruction, requiring
 at least one
$b$-tagged jet and $b\bar{b}$ mass smaller than 100 GeV$/c^2$. The second
region probes $b\bar{b}$ and $W^+W^-b\bar{b}$ mass reconstruction, requiring
 at least one
$b$-tagged jet and $W^\pm b\bar{b}$ mass smaller than 250 GeV$/c^2$. The
third region tests the modeling of $W^\pm b\bar{b}$ and $b\bar{b}$  mass
reconstruction, requiring at least one $b$-tagged jet and
$W^+W^-b\bar{b}$ mass less than 450 GeV$/c^2$. The fourth region tests
the modeling of the $W$-boson-plus-jets background, requiring exactly
zero $b$-tagged jets and $W^+W^-b\bar{b}$ mass greater than 450 GeV$/c^2$.  Assuming an $H^0$ production cross section of 250
fb, each control region is expected to have negligible signal
contamination, with the exception of the zero $b$-tag region which
would include signal events at approximately 10\% of the sample. For
two of the control regions, Fig.~\ref{fig:ctrl_region}  shows the
reconstructed $b\bar{b}$ mass distributions which, along with other
similar distributions, indicate that the background mass distributions
are well modeled within systematic uncertainties.

\begin{figure}[!h]
\includegraphics[width=0.9\linewidth]{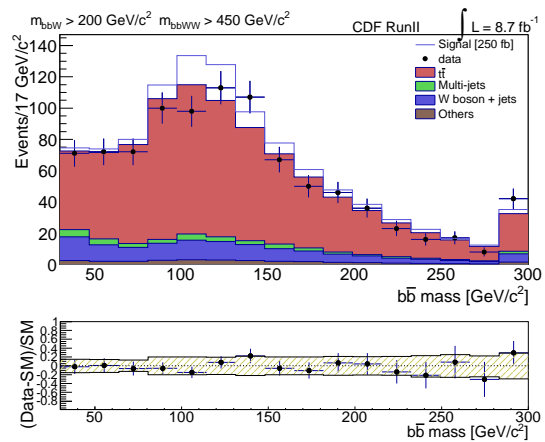}\\
\caption{Distribution of events versus reconstructed $b\bar{b}$  invariant mass ($m_{b\bar{b}}$), for observed data and expected backgrounds in the signal region.  A signal
  hypothesis is shown, assuming a total cross section of 250 fb, 
  $m_{H^0}=500$ GeV$/c^2$, and $m_{H^\pm}=300$ GeV$/c^2$.  See
  Fig~\ref{fig:ctrl_region} for descriptions of lower panel and hatching.}
\label{fig:sig_region}
\end{figure}

\begin{figure}[!h]
\includegraphics[width=0.7\linewidth]{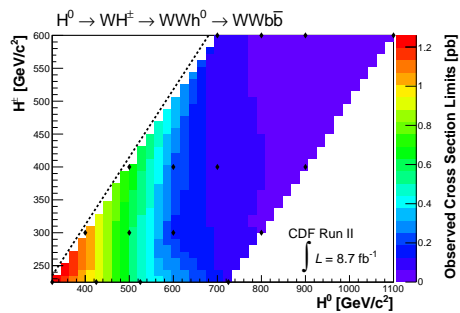}\\
\caption{  Upper limits at 95\% C.L.
  on the cross section times branching fraction as a function of the Higgs-boson masses $m_{H^\pm}$
  and $m_{H^0}$; $m_{h^0}$ is fixed to 126 GeV$/c^2$ in each case. Diamonds show
  the grid of probed masses; the intermediate values are
  interpolated. }
\label{fig:limits}
\end{figure}

Figure~\ref{fig:sig_region} shows the observed distribution of events
in a representative signal region 
compared to possible signals and estimated backgrounds.  At each Higgs-boson
mass hypothesis, we fit the
most likely value of the Higgs-boson
 cross section by performing a
 maximum-likelihood fit in the binned $m_{b\bar{b}}$ distribution, allowing for
systematic and statistical fluctuations via template
morphing~\cite{morph}.  No evidence is found for the presence of Higgs-boson cascade decays
in $WWb\bar{b}$ events. We set upper limits on Higgs  production
 at 95\% confidence level
 using the CLs method~\cite{MCLimit}, without profiling the systematic uncertainties.
 The observed limits are
 consistent with expectation for the background-only hypothesis.  See Fig.~\ref{fig:limits} and Table~\ref{tab:lim}.

\begin{table}[h!]
\caption{ Signal region definitions and  expected and observed
   95\% C.L. upper limits on the production cross section times branching
  fraction for each Higgs-boson mass hypothesis. Theoretical predictions are also shown~\cite{theory1, theory2,theory}.}
\label{tab:lim}
\centering
\begin{tabular}{l|r@{\hspace{2mm}}|r@{\hspace{2mm}}|@{\hspace{6mm}}cc}
\hline\hline
$(m_{H^0},m_{H^\pm})$  & $m_{H^\pm}$ & $m_{H^0}$  & Exp (Obs) & Theory
\\
{\tiny (GeV$/c^2$)}  &{\tiny  (GeV$/c^2$)} &{\tiny  (GeV$/c^2$)} & Limit (fb)  & (fb) \\
\hline
325, 225		& $>175	$ & $>275$ & $1100$ $(1300)$ &$34$\\
400, 300		& $>225	$ & $>325$ & $960$ $ (1100)$ &$18$\\ 
425, 225		& $>200	$ & $>375$ & $900$ $ (960)$ &$13$\\
500, 300		& $>200	$ & $>450$ & $470$ $ (590)$ &$3.9$\\
500, 400		& $>350	$ & $>450$ & $510$ $ (700)$ &$3.9$\\
525, 225		& $>100	$ & $>500$ & $420$ $ (460)$ &$2.5$\\
600, 300		& $>200	$ & $>550$ & $200$ $ (180)$ &$0.76$\\
600, 400		& $>350	$ & $>550$ & $210$ $ (250)$&$0.76$\\
700, 400		& $>325	$ & $>650$ & $90$ $ (100)$ &$0.15$\\
700, 600		& $>450	$ & $>650$ & $10$ $ (96)$&$0.15$\\
725, 225		& $>425 $ & $>700$ & $90$ $ (120)$ &$0.10$\\
800, 300		& $>275	$ & $>750$ & $50$ $ (51)$ & $3\times 10^{-2}$\\
800, 600		& $>475	$ & $>725$ & $43$ $ (46)$  &
$3\times 10^{-2}$\\
900, 400		& $>450	$ & $>775$ & $28$ $ (36)$ &
$6\times 10^{-3}$\\
900, 600		& $>475	$ & $>800$ & $24$ $ (29)$ &
$6\times 10^{-3}$ \\
1100, 600		& $>475	$ & $>975$ & $13$ $ (15)$ &
$2\times 10^{-4}$ \\
\hline\hline
\end{tabular}
\end{table}

In conclusion, we report on the first search for multiple Higgs bosons
in cascade decays.
 For each accepted event, we reconstruct the 
lightest neutral Higgs-boson mass ($m_{b\bar{b}}$), and find the CDF
data to be consistent with standard model
background predictions. We calculate 95\% C.L. upper limits  on
the cross section of such Higgs-boson production, assuming 100\%
branching ratio of $H^0$ to $W^\pm H^\mp$ and $H^\pm$ to $W^\pm h^0$,
from $1.3$ pb to $0.015$ pb
 for  masses ranging from $(m_{H^0}=325,m_{H^\pm}=225)$ GeV$/c^2$ to
 $(m_{H^0}=1100,m_{H^\pm}=600)$ GeV$/c^2$ respectively, and
 interpret the limits in terms of a simplified
two-Higgs-doublet model.  While the limits cited here do not exclude any region in the $m_{H^0}-m_{H^\pm}$-plane in the simplified model used, there are the first such limits available. The larger
center-of-mass energy and integrated luminosity of data collected by
the LHC experiments are likely to have the sensitivity to discover or exclude such models.

We thank the Fermilab staff and the technical staffs of the participating institutions for their vital contributions. This work was supported by the U.S. Department of Energy and National Science Foundation; the Italian Istituto Nazionale di Fisica Nucleare; the Ministry of Education, Culture, Sports, Science and Technology of Japan; the Natural Sciences and Engineering Research Council of Canada; the National Science Council of the Republic of China; the Swiss National Science Foundation; the A.P. Sloan Foundation; the Bundesministerium f\"ur Bildung und Forschung, Germany; the Korean World Class University Program, the National Research Foundation of Korea; the Science and Technology Facilities Council and the Royal Society, UK; the Russian Foundation for Basic Research; the Ministerio de Ciencia e Innovaci\'{o}n, and Programa Consolider-Ingenio 2010, Spain; the Slovak R\&D Agency; the Academy of Finland; and the Australian Research Council (ARC).


\begin{thebibliography}{99}
\bibitem{coordinates}
CDF uses a cylindrical coordinate system with
  the $z$ axis along the proton beam axis. For a particle or a jet, pseudorapidity is $\eta
  \equiv -\ln(\tan (\theta/2))$, where $\theta$ is the polar angle of
  a particle or a jet
  relative to the proton beam direction, and $\phi$ is the azimuthal
  angle  while transverse momentum is $p_T = |p|\ \sin\theta$, and the
  transverse energy is $E_T = E\ \sin\theta$.
\bibitem{metdef} 
Missing transverse momentum, $\missET$, is defined as the magnitude of
the vector $ -\sum_i E_T^i \vec n_i$ where $E_T^i$ are the magnitudes
of transverse energy contained in each calorimeter tower $i$, and
$\vec n_i$ is the unit vector from the interaction vertex to the tower
cetroid in the transverse ($x,y$) plane.
\end{thebibliography}

\end{document}